\documentstyle[preprint,aps]{revtex}

\def\r{\rho}

\def\aa{{\alpha,\alpha'}}
\def\kk{{k,k'}}
\def\x{{\bf x}}
\def\y{{\bf y}}
\begin{document}
\draft

\title{N--d scattering above the deuteron breakup threshold}

\author{ A.~Kievsky$^1$, M. Viviani$^1$ and S.~Rosati$^{1,2}$}
\address{ $^1$Istituto Nazionale di Fisica Nucleare, Piazza Torricelli 2,
          56100 Pisa, Italy }
\address{ $^2$Dipartimento di Fisica, Universita' di Pisa, Piazza Torricelli 2,
          56100 Pisa, Italy }

\date{\today}

\maketitle

\abstract{ The complex Kohn variational principle and the (correlated)
Hyperspherical Harmonics  technique are applied to study the
N--d scattering above the deuteron breakup threshold. The 
configuration with three outgoing nucleons is explicitly taken
into account by solving a set of differential equations with outgoing
boundary conditions.  
A convenient procedure is used to obtain the correct boundary
conditions at values of the hyperradius $\approx 100$ fm. The
inclusion of the Coulomb potential is straightforward and does not
give additional difficulties. Numerical results have been obtained
for a simple s-wave central potential. They are in nice agreement with
the benchmarks produced by different groups using the Faddeev
technique. Comparisons are also done with experimental 
elastic N--d cross section at several energies.  
}

\narrowtext
\bigskip

\noindent{PACS numbers: 25.10+s, 03.65.Nk, 25.55.Ci}

\bigskip
\noindent{key words: N-d scattering, Kohn variational principle, phase shift,
          breakup, elastic cross section, inelastic cross section}

\newpage

One of the main objectives in nuclear physics is the knowledge of 
the nuclear interaction. In practice, the
two-nucleon scattering data are used to determine the on-shell nucleon-nucleon 
(NN) interaction. The off-shell properties of nuclear potentials 
and many-body force contributions must be tested in systems with $A>2$.
As a consequence, much work has to be devoted to the understanding of the 
three nucleon bound and scattering states. The Faddeev theory has been 
extensively applied to this problem, and, in particular, to the study  
of  the n-d scattering under and above the deuteron breakup threshold.
The corresponding Faddeev equations
in momentum space were originally solved by  Kloet and Tjon~\cite{Kloet} 
for a central s-wave potential. At present, different
numerical techniques are available to solve the Faddeev equations in
configuration and momentum space. 
In refs.~\cite{Bench1,Bench2} benchmark calculations for n-d scattering
were given as reference for new techniques. Realistic potentials
have been used to calculate the n-d scattering cross--section
at different energies~\cite{Gloec}. For the p-d channel 
many accurate experimental data are available, but 
the Faddeev approach to this process becomes difficult due to the Coulomb
repulsion~\cite{Alt,Merkuriev,Vanvan}. 
Such a difficulty is not present in the 
variational technique developed in ref.~\cite{KVR94}
for the N-d scattering below the deuteron breakup. 
The extension of this
above the deuteron breakup threshold is the object of the present paper.
In particular, the asymptotic conditions
to describe three outgoing nucleons in the n-n-p and p-p-n states
are explicitly taken into account.

In the following the important aspects of the approach are briefly
outlined and various results obtained for a simple s-wave central
potential are reported. More details on the adopted procedures 
and results for realistic potentials 
will be presented in a forthcoming paper~\cite{next}.

Following ref.~\cite{KVR94}, the wave function of the system is written as a
sum of two terms
\begin{equation}
  \Psi = \Psi_C + \Psi_A\ .
\end{equation}
The $\Psi_A$ term is  solution of the
Schroedinger equation in the asymptotic region where the incident nucleon
and the deuteron are well apart.
The $\Psi_C$ term must guarantee an accurate description of the 
system when the three--nucleons are close to each other; moreover,
for large interparticle separations it has to describe the breakup 
configurations. $\Psi_C$ is written as a sum of  
channel contributions, labelled by the angular--spin--isospin quantum 
number. The associated 
two--dimensional spatial amplitudes are expanded in terms of 
the Pair Correlated Hyperspherical Harmonic (PHH) basis~\cite{KVR93}.

The wave function corresponding to an asymptotic state
${}^{(2S+1)}L_J$ has the form 

\begin{equation}
 \Psi_{LSJ}=\sum_{i=1,3}\left[ \Psi_C(\x_i,\y_i) +
 \Omega^{in}_{LSJ}(\x_i,\y_i) - \sum_{L'S'}
 {}^J{\cal S}^{SS'}_{LL'} \Omega^{out}_{L'S'J}(\x_i,\y_i)\right]
 \ , 
\end{equation}
where the summation is extended over the three different choices of the 
Jacobi coordinates $(\x ,\y)$. 
$L$ is the relative angular momentum between the incident nucleon and the 
deuteron, $S$ is the spin obtained by coupling the spin $j= 1$ of the deuteron 
with the spin 1/2 of the third nucleon and $J$ is the total angular 
momentum of the system.
${}^J{\cal S}^{SS'}_{LL'}$ are the collision
matrix ($\cal S$--matrix) elements describing the $2\rightarrow 2$
elastic scattering.
The function $\Omega^\lambda_{LSJ}$ is
the ingoing ($\lambda\equiv in$) or the outgoing ($\lambda\equiv out$) 
solution of the two-body N-d Schroedinger equation in the asymptotic region.
These solutions contain suitable  regularizing  factors 
at small distances.
The explicit form of $\Psi_C$ is
\begin{eqnarray}
     \Psi_C(\x_i,\y_i) &= &\sum_\alpha \phi_\alpha(x_i,y_i) 
     {\cal Y}_\alpha (jk,i)\ ,  \\ 
     {\cal Y}_\alpha (jk,i) &= & \Bigl\{\bigl[ Y_{\ell_\alpha}(\hat x_i)  
     Y_{L_\alpha}(\hat y_i) 
     \bigr]_{\Lambda_\alpha} \bigl [ s_\alpha^{jk} s_\alpha^i \bigr ]
     _{S_\alpha} \Bigr \}_{J J_z} \; 
     \bigl [ t_\alpha^{jk} t_\alpha^i \bigr ]_{T T_z}\ , \\
     \phi_\alpha(x_i,y_i) &=& \rho^{\ell_\alpha+L_\alpha} f_\alpha (x_i)
     \left[ \sum_k u^\alpha_k(\rho) {}^{(2)}P^{\ell_\alpha,L_\alpha}_k(\phi_i)
     \right] \ ,
\end{eqnarray}
where the hyperspherical variables $\rho^2=x_i^2+y_i^2$ and 
$\cos\phi_i= x_i/\rho$ have been introduced. The pair correlation functions
$f_\alpha$ are solutions of a two--body Schroedinger--like
equation~\cite{KVR93} and are included to accelerate the convergence of
the expansion. 
At small interparticle distances they take into account the correlations
introduced by the strong repulsion of the NN potential and  go smoothly
to unity as the interparticle distance increases.
The index $\alpha$ in the sum 
runs over all the channels compatible with the total angular momentum $J$ 
value, and the antisymmetrization and parity conditions. In numerical
applications  
the sum is truncated after including all the important channels.
The quantities to be determined in the wave function specified by eqs.(2-5)
are the hyperradial functions 
$u^\alpha_k(\rho)$ and the collision matrix elements. They will be 
calculated by means of the Kohn variational principle. 

Below the deuteron breakup the elastic collision matrix is unitary and the
problem can be also formulated in terms of the reactance matrix, as 
done in \cite{KVR94}.
Above the deuteron breakup the complex form of the Kohn variational 
principle~\cite{CKohn} is better suited to describe three outgoing particles.
In this case, the  ${\cal S}$--matrix and the hyperradial
functions are determined through the stationary value of the functional 
\begin{equation}
 [^J{\cal S}^{SS}_{LL}]={} ^J{\cal S}^{SS}_{LL} + 
                       i<\Psi^*_{LSJ}|H-E|\Psi_{LSJ}> \ .
 \label{eq:ckohn}
\end {equation}

Let us first point out the main points of the procedure.
After performing the variation with respect to
the hyperradial functions, the following set of coupled equations is
obtained:
\begin{equation}
  \sum_{\alpha',k'}
       \Bigl[ A^\aa_\kk (\r ){d^2\over d\r^2}+ B^\aa_\kk (\r ){d\over d\r}
        + C^\aa_\kk (\r )+{m\over\hbar^2} E\; N^\aa_\kk (\r )\Bigr ]  
        u^{\alpha'}_{k'}(\r)= D^\lambda_{\alpha k}(\rho) \ .
   \label{eq:siste}
\end{equation}
Details  on the explicit form of the coefficients $A,B,C,N$ 
and the inhomogeneous
term $D$ can be found in refs.~\cite{KVR94,KVR93}. 
For each asymptotic state $^{(2S+1)}L_J$  two different inhomogeneous terms
can be constructed in correspondence to the
asymptotic $\Omega^\lambda_{LSJ}$ functions with
$\lambda\equiv in$ or $out$. Two different sets of
hyperradial functions are then obtained by solving 
the system of eqs.(7) for the two choices of $\lambda$.
In the subsequent step the two sets are combined
to minimize the functional $[{}^J{\cal S}^{SS}_{LL}]$ with respect to
variations of the ${\cal 
S}$--matrix elements. This is the first order solution; the second order
estimate is calculated by replacing the first order solution in
eq.~(\ref{eq:ckohn}). 

Appropriate asymptotic conditions must be imposed on the hyperradial
functions $u^\alpha_k(\rho)$ to completely determine the problem.
When $\rho\rightarrow \infty$, they should vanish below the deuteron
breakup,  whereas,  for positive total energy $E$,
they should be proportional to $\exp(i {\sqrt E}\rho)$.
However, this behavior is reached only for very large values of the
hyperradius~\cite{GP92}, and, in addition, 
the presence of the Coulomb potential modifies the free outgoing
wave also at infinity.
Therefore, it is convenient to use the asymptotic behavior
of the coefficients $A$, $B$, $C$, $N$ and $D$, entering 
eqs.~(\ref{eq:siste}), to obtain the solutions at large but finite $\rho$.
Neglecting terms going to zero faster than $\rho^{-3}$, the asymptotic
expression for the set of differential equations can be cast in the
form 
\begin{equation}
  \sum_{\alpha',k'} \Bigl[ \delta_\aa\delta_\kk 
 ({d^2\over d\r^2}-{ {\cal L}({\cal L}+1)\over\r^2} + Q^2)  
 - {2\; Q\; \chi^\aa_\kk \over\r }+{h^\aa_\kk \over \rho^3}\Bigr ]  
              U_{\alpha' k'}(\r)= 0 \ .
\label{eq:siste1}
\end{equation}
The $\chi$ term originates from the Coulomb potential matrix elements
and it shows the expected $1/ \rho$ behaviour. The kinetic energy
operator and the nuclear potential contribute both to the $h$ term.
The final form~(\ref{eq:siste1}) is obtained after orthonomalizing
the PHH states at $\rho = \infty$. In the above equation,
$Q^2=mE/\hbar^2$, ${\cal L}=\ell_\alpha +L_\alpha + 2k +3/2$ and
$U_{\alpha k}(\rho)$ are linear combinations of the functions
$\rho^{\ell_\alpha+L_\alpha-5/2}u^\alpha_k(\rho)$. The total number of
coupled equations in eqs.~(\ref{eq:siste}) and ~(\ref{eq:siste1}) is $N_{eq}$,
corresponding to all the considered values of $\alpha,k$.

For n-d scattering the $\chi$ term is zero and,  if the coupling
term $h$ is neglected, the outgoing solutions of (\ref{eq:siste1}) are the
Hankel  functions $H^{(1)}(Q\rho)$. In order to take into account the
coupling terms, 
$N_{eq}$ different solutions of eqs.~(\ref{eq:siste1}) of the kind
\begin{equation}
   W^{(\alpha_0 k_0)}_{\alpha k}(\rho)= 
   \sum_{\alpha_0',k_0'}
   \sum_{n=0,1,2,\ldots} 
   {\Gamma^{(\alpha_0' k_0')}_{\alpha k}(n) \over \rho^n}
   \left (    e^{-i \chi\log2 Q\rho} \right)_{\alpha_0',\alpha_0}
    ^{k_0',k_0} e^{i Q\rho}   \ ,
  \label{eq:cubo}
\end{equation}
where $\chi$ is the matrix entering eq.~(\ref{eq:siste1}),
are obtained by choosing $\Gamma^{(\alpha_0,k_0)}_{\alpha
k}(n=0)=\delta_{\alpha\alpha_0} \delta_{k k_0}$.
The $n>0$ coefficients $\Gamma$ are
determined by recurrence relations obtained from
eq.~(\ref{eq:siste1}), as done, for example, in
ref.~\cite{Mandel}. 

The solutions of eqs.~(\ref{eq:siste}) are then matched to 
specific superpositions of the functions $W^{(\alpha_0 k_0)}_{\alpha
k}(\rho)$ by imposing the continuity of the logarithmic derivative at a
given value $\rho=\rho_0$.  The value of the matching radius $\rho_0$ is not
relevant, provided that the asymptotic form~(\ref{eq:siste1}) 
is reached, which is rather well verified for
$\rho_0>80$ fm. With such a condition, it has been numerically tested
that the  solutions are insensitive to variation of  $\rho_0$, 
even in the presence of Coulomb potential terms.
For $\rho\rightarrow \infty$,  such solutions evolve as
\begin{equation}
  U_{\alpha k}(\rho) \rightarrow
   - \sum_{\alpha_0 k_0} 
  \left ( e^{-i \chi\log2 Q\rho} \right)_{\alpha,\alpha_0}^{k,k_0}
   S_{\alpha_0 k_0} e^{i Q\rho} \ ,
    \label{eq:asy2}
\end{equation}
corresponding to the correct asymptotic behavior of three outgoing
particles interacting via long--range Coulomb potentials~\cite{Merkuriev}.
From the above equation, it results  that the $S_{\alpha k}$ parameters
are just the inelastic ${\cal S}$--matrix elements describing the
$2\rightarrow3$ breakup process. In the n-d case, 
the hyperradial functions asymptotically reduce
to $U_{\alpha k}(\rho)\rightarrow - S_{\alpha k} e^{i Q\rho}$.

The NN interaction model considered in the present paper 
is the $s$--wave potential of Malfliet and Tjon I-III~\cite{MT}, with
the parameter values given in ref.~\cite{Bench2}. Correspondingly, for
n-d scattering, the function $\Psi_C({\bf x},{\bf y})$ defined in eqs.~(3--5)
includes only  $\ell_\alpha=0$ channels. Therefore,
the total number of channels for the states with total spin $S={1\over
2}$ ($S={3\over 2}$)  is simply two (one). 
On the other hand, the Coulomb potential is active in all the waves, and
so, in the p-d case, also channels with angular momenta $\ell_\alpha$
larger than zero have been included. 
For central potentials  the
elastic part of the collision matrix  does not 
depend  on $J$ and, moreover,  $S=S'$, $L=L'$. Thus, 
${}^J{\cal S}^{SS}_{LL}$ has been expressed in the
usual form ${}^{2S+1}\eta_L\; \exp(2 i \; {}^{2S+1}\delta_L)$.

The number of hyperspherical states, i.e.
of hyperradial functions, included in each
channel has been increased until the convergence is reached. 
Typically, eight hyperradial functions per channel are enough for a four
digit accuracy in the phase shift parameters.
The rate of convergence for the
$s$--wave phase--shifts $^2\delta_0$ and $^4\delta_0$ and
inelasticity parameters $^2\eta_0$ and $^4\eta_0$ in function
of the number $N_\alpha$ of considered states (equations) per channel
$\alpha$  is shown in table 1 at $E_N=14.1$ MeV.
A similar trend is found at $E_N=42$ MeV and the corresponding
converged results are shown in table 2. For n-d scattering
a direct comparison with the benchmark results~\cite{Bench2} can be done,
showing a very good agreement. 
The phase shift and inelasticity parameters have been also calculated
for p-d scattering. Convergence patterns similar to the n-d case are obtained,
as can be seen in table 1. It can be observed that there is
a constant difference of about $3^\circ$ between the 
p-d and n-d $s$--wave phase--shifts at the energy values considered.
This difference is in reasonable agreement with the corresponding results
obtained by other authors~\cite{Alt,Merkuriev}.

To compare with experimental
results, the elastic cross section has been calculated
for several energy values $E_N$ of the incident nucleon. Since the
nuclear potential  here considered is only 
$s$--wave active, the differential cross section starts 
deviating from the experimental points above $E_N=20$ MeV~\cite{Kloet}.
For lower $E_N$ values, partial waves up to $L=8$  give sizeable
contributions and have been included.
The calculated n-d elastic cross section is presented in fig.1 
at neutron energies $E_n=3,9,18$ MeV, together with the experimental data of 
ref.~\cite{Schwarz}. The p-d elastic cross section is given in fig.2 at 
proton energies $E_p=3,6,9,18$ MeV, 
in conjunction with the high precision data of ref.~\cite{Sagara}.
Despite of its simple form, the MT(I-III) potential
reproduces in a reasonable way the experimental cross sections. 
The differences found at small angles disappear when more realistic 
NN interactions are used~\cite{KRTV96}.

The calculation of the breakup cross section is easily performed once
the coefficients $S_{\alpha k}$ given in eq.~(\ref{eq:asy2}) are
known. The ${\cal S}$--matrix unitarity imposes the following relation between 
the elastic and inelastic parameters
\begin{equation}
  |{}^{2S+1}\eta_L|^2+\sum_{\alpha k} |S_{\alpha k}|^2 = 1 \ .
  \label{eq:unit}
\end{equation}
In all the cases we have  considered, this relation 
is well verified numerically with a precision of $10^{-5}$. An example
of n-d and p-d breakup cross sections, for the space star configuration, is
reported in fig.3. For the sake of comparison, the n-d breakup cross section
calculated by means of the Faddeev technique~\cite{Bench2} is
reported, as well. From inspection of the fig.3, it
can be noticed the good agreement between the results of the two
n-d calculations. 

The inelasticity parameter ${}^{2S+1}\eta_L$ goes  to one as
$L$ increases, as, for example,  it has been found in
refs.~\cite{Kloet,Gloec} for n-d scattering. 
For instance, at $E_n=14.1$ MeV, $(1-{}^{2S+1}\eta_L)<  10^{-3}$  already for
$L\ge 4$. A similar behavior has been found for the p-d case.
The contributions to both elastic and inelastic cross sections
of channels with $\ell_\alpha> 0$, included in the p-d case,
have been found to be nearly  negligible.

We conclude with a few remarks. First of all,
the Kohn variational principle has been successfully applied to treat
scattering processes above the deuteron breakup threshold. The complex form
of the principle is well suited to take into account the boundary
conditions for the three  outgoing nucleons and
to obtain the second order estimate of the ${\cal S}$--matrix.
The expansion of the wave function in the PHH basis allows for
lowering the number of hyperspherical states to be included. The problem
reduces to the solution of a set of second--order inhomogenous
differential equations with outgoing boundary conditions.
For p-d scattering the equations are coupled
even in the asymptotic region due to the Coulomb potential. However,
a simple technique can be used to calculate the proper boundary
conditions for the wave function at values of the hyperradius
$\rho\approx 100$ fm.  The results obtained for the complex $s$--wave
n--d phase--shift  parameters are in complete agreement with the 
benchmark calculations of ref.~\cite{Bench2}. The elastic cross
sections have been compared with the experimental data. For both
processes, n-d and p-d, an overall good agreement has been observed,
with small differences that may be ascribed to the rather simple
potential used.  

The extension of the method to realistic NN interactions will be the
subject of a subsequent paper.



\begin{table}
\caption{
Real part of the $s$--wave phase--shift $^{2S+1}\delta_0$ 
(in degrees) and inelasticity parameter $^{2S+1}\eta_0$ for the 
doublet and quartet spin states are given as a function of the number 
$N_\alpha$ of hyperspherical functions considered per channel.
The incident nucleon energy is $E_N=14.1$ MeV.}
\label{tab:phasec}
\begin{tabular}{c|cc|cc}
 $N_\alpha$    & $^2\delta_0$ & $^2\eta_0$ & $^4\delta_0$ & $^4\eta_0$ \\ 
\hline
 & \multicolumn{4}{c}{n-d } \\
\hline
  2        &  97.96       & 0.5093     &  67.01       &  0.9933    \\
  4        & 105.47       & 0.4652     &  68.88       &  0.9788    \\
  6        & 105.51       & 0.4650     &  68.94       &  0.9784    \\
  8        & 105.50       & 0.4649     &  68.95       &  0.9782    \\
\hline
 & \multicolumn{4}{c}{p-d } \\
\hline
  2        & 101.16       & 0.5430     &  70.92       &  0.9905    \\
  4        & 108.41       & 0.4989     &  72.53       &  0.9801    \\
  6        & 108.44       & 0.4986     &  72.59       &  0.9797    \\
  8        & 108.43       & 0.4985     &  72.60       &  0.9795    \\
\hline
\end{tabular}
\end{table}

\begin{table}
\caption{Results obtained for the
real part of the $s$--wave phase--shift $^{2S+1}\delta_0$ 
(in degrees) and the inelasticity parameter $^{2S+1}\eta_0$ 
are given at the specified incident nucleon energies. 
For the n-d process, the benchmarks results of ref.~\protect\cite{Bench2}
are reported for the sake of comparison.}
\label{tab:phase}
\begin{tabular}{c|cc|cc}
 & \multicolumn{4}{c}{n-d at $E_n=14.1$ MeV} \\
\hline
           & $^2\delta_0$ & $^2\eta_0$ & $^4\delta_0$ & $^4\eta_0$ \\ 
\hline
present    & 105.50       & 0.4649     &  68.95       &  0.9782    \\
Los Alamos
           & 105.48       & 0.4648     &  68.95       &  0.9782    \\
Bochum
           & 105.50       & 0.4649     &  68.96       &  0.9782    \\
\hline
 & \multicolumn{4}{c}{n-d at $ E_n=42.0$ MeV} \\
\hline
present    &  41.33       & 0.5026     &  37.71       &  0.9034    \\
Los Alamos
           &  41.34       & 0.5024     &  37.71       &  0.9035    \\
Bochum
           &  41.37       & 0.5022     &  37.71       &  0.9033    \\
\hline
 & \multicolumn{4}{c}{p-d at $E_p=14.1$ MeV} \\
\hline
present    & 108.43       & 0.4985     &  72.60       &  0.9795    \\
\hline
 & \multicolumn{4}{c}{p-d at $E_p=42.0$ MeV} \\
\hline
present    &  43.65       & 0.5058     &  39.94       &  0.9047    \\
\hline
\end{tabular}
\end{table}



\begin{figure}
\caption{Elastic n-d cross section at  neutron energies $E_n=3,9,18$ MeV.
The experimental data are from ref.~\protect\cite{Schwarz}}
\label{fig:n-d}
\end{figure}

\begin{figure}
\caption{Elastic p-d cross section at proton energies $E_p=3,6,9,18$ MeV.
The experimental data are from ref.~\protect\cite{Sagara}}
\label{fig:p-d}
\end{figure}

\begin{figure}
\caption{Laboratory breakup cross section for n-d and p-d scattering
at $E_N=14.1$ MeV versus the arclength $S$, for the configuration
where two neutrons or two protons are detected at angles
$\theta_1=51.02$, $\theta_2=51.02$ and $\varphi=120$ degrees. For the
sake of comparison with the calculation of ref.~\protect\cite{Bench2},
only the contributions from s-waves have been included.
The solid (dotted) curve corresponds to n-d (p-d) scattering. The
solid dots are the n-d results taken from ref.~\protect\cite{Bench2}.
}
\end{figure}

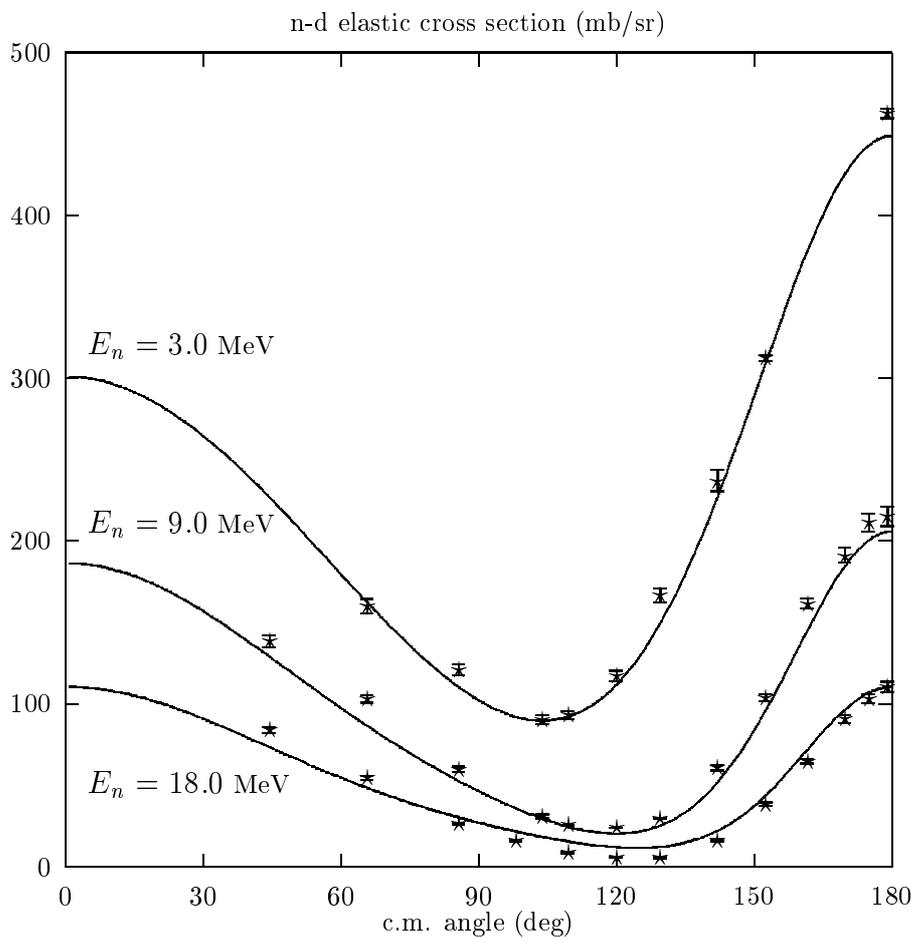
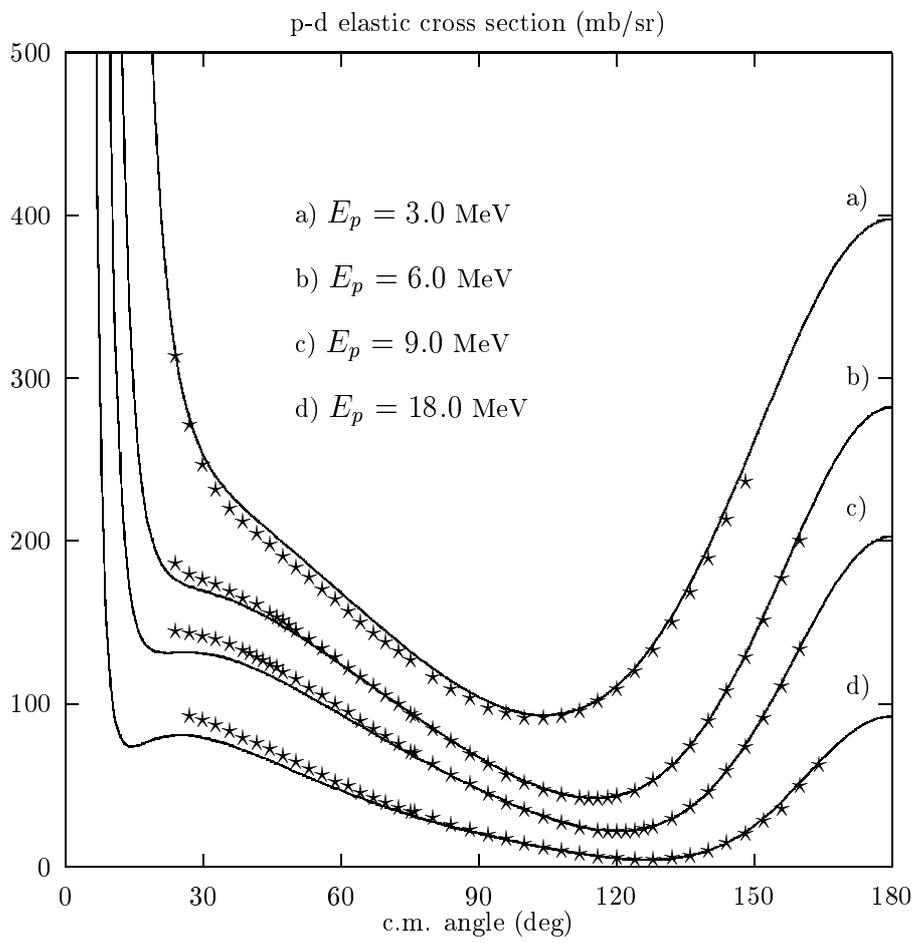
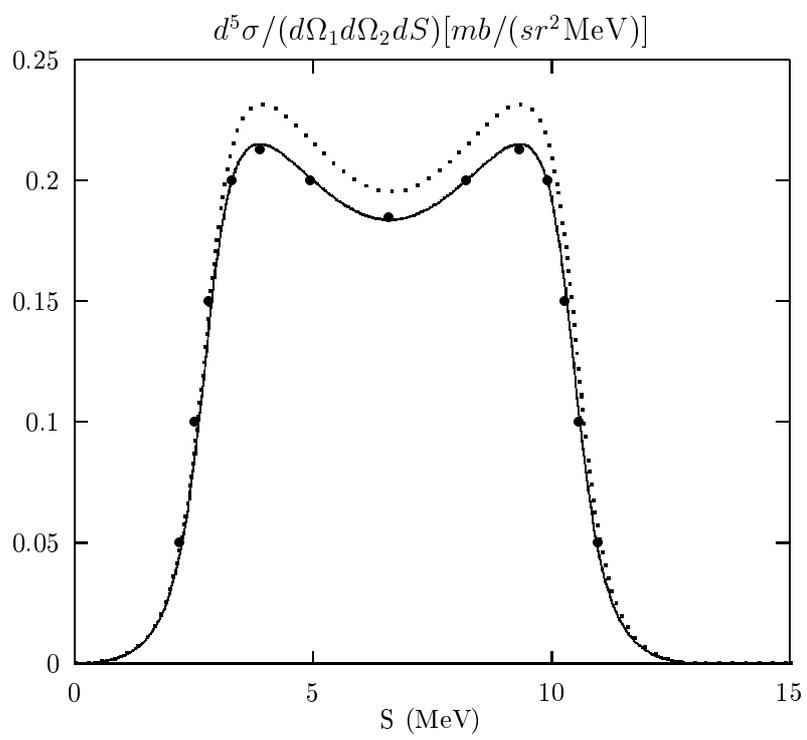
\end{document}